# Direction-Selective Wave Freezing and Amplification at a Hyperbolic Time Interface

ZHAO WANG,[1] AI GANG,[2] XINGHONG ZHU,[1] HONGRU MA,[2] WEN XIAO[2,*] HUANYANG CHEN[1,†]

[1] Department of Physics, Xiamen University, Xiamen 361005, China

[2] College of Science, Shantou University, Shantou, 515063, China

[*] Corresponding author: xiaowen@stu.edu.cn

[†] Corresponding author: kenyon@xmu.edu.cn

**Abstract**: Hyperbolic media are well known for converting high k-components that are evanescent in conventional dielectrics into propagating bulk waves through their open equifrequency contours. Here, we reveal a complementary temporal effect: after a sudden transition into an effectively nondispersive hyperbolic state, conservation of the full wavevector causes the indefinite dispersion to partition momentum space into real-, zero-, and purely imaginary-frequency regimes. Consequently, p-polarized waves undergo conventional temporal scattering, critical magnetic-field freezing, or exponential growth and decay, depending solely on their conserved wavevector direction, whereas s-polarized waves remain in the real-frequency regime. A second temporal boundary releases the frozen or amplified fields into propagating waves at original frequency. Analytical temporal boundary theory, k-space pulse reconstruction, and finite-difference time-domain simulations corroborate these dynamics. These results establish hyperbolic temporal boundaries as a compact route to direction-selective imaginary frequency dynamics and wave amplification without Floquet periodicity.

## 1.Introduction:

Temporal boundaries have attracted significant research interest due to their novel physical properties, including frequency conversion, momentum conservation, and energy amplification [1–4]. In contrast to conventional spatial boundaries where electromagnetic parameters vary in space, a temporal boundary involves a global, abrupt change of material properties (e.g., permittivity or conductivity) at a specific moment [5–11]. When the permittivity abruptly changes within a timescale shorter than the optical oscillation period, the incident wave splits into a forward-propagating (time-refracted) wave and a backward-propagating (time-reflected) wave, serving as the time-domain analogue to conventional spatial reflection and refraction [3,6]. This temporal degree of freedom provides a new paradigm to control electromagnetic waves and lead to unprecedented optical phenomena such as the inverse prism [5], nonreciprocal [12,13], temporal aiming [14], polarization conversion [13,15] and non-resonant lasing [16,17].

Among the phenomena enabled by temporal modulation, complex-frequency dynamics are particularly interesting as they govern the transition from stable oscillation to exponential temporal evolution, which provide a temporal counterpart to spatial evanescence. An imaginary wavevector component produces exponential variation in space, while a complex frequency gives rise to temporally growing and decaying modes. In photonic time crystals, such modes occupy momentum bandgaps and underpin non-resonant amplification and proposed lasing by drawing energy from periodic modulation [16–21]. The canonical isotropic photonic time crystal is selective in wave-number magnitude, but not in wave-vector direction or polarization, because all directions sharing the same $|k|$ possess the same Floquet spectrum. Anisotropic photonic time crystals can reshape this momentum-space response and have enabled direction-dependent radiation and selective amplification or attenuation of nonuniform waves [17,22]; however, their complex-frequency states still originate from repeated Floquet scattering under periodic modulation. At the single-boundary limit, a sudden transition to an isotropic non-Foster negative-permittivity state can produce non-oscillatory growing and decaying waves without temporal periodicity [23], yet the response remains intrinsically degenerate with respect to propagation direction and polarization. These developments raise a fundamental question: can temporal evanescence be sculpted at a single temporal boundary, so that only selected polarizations and wave-vector directions enter an imaginary-frequency regime while neighboring channels remain oscillatory?

Hyperbolic dispersion offers a natural route to such single boundary momentum-space control. Characterized by diagonal components of the permittivity tensor with opposite signs, hyperbolic media possess open equifrequency contours and a sign-indefinite dispersion relation [24–27]. At a spatial interface, this unusual dispersion reshapes the boundary between propagating and evanescent states, allowing high k-components that decay in conventional dielectrics to propagate as bulk modes in hyperbolic medium. This spatial behavior naturally suggests a counterpart at a temporal boundary. As spatial homogeneity conserves the full wavevector across a temporal transition, the converted frequency is determined by the temporal Snell's equations. Therefore, the indefinite dispersion of a hyperbolic medium provides a potential means of making the onset of non-oscillatory temporal dynamics depend on both polarization and wavevector direction. Rather than producing a direction-degenerate imaginary-frequency response, as in an isotropic negative-permittivity transition, a hyperbolic temporal boundary may therefore enable the imaginary frequency region itself to be sculpted in momentum space.

In this work, we show that an abrupt dielectric-to-hyperbolic transition creates a polarization- and direction-selective temporal stability landscape at a single temporal boundary. For p-polarized waves, different conserved wavevector directions access oscillatory real-frequency states, a zero-frequency critical state, or purely imaginary-frequency states, whereas s-polarized waves remain in the real-frequency regime. The purely imaginary-frequency sector constitutes a direction-selective form of temporal evanescence: magnetic-field freezing occurs at its critical boundary, while exponential amplification arises from its growing temporal branch. A second temporal transition back to the initial dielectric converts the frozen or amplified fields into propagating waves at the original frequency. Our results establish hyperbolic temporal boundaries as a Floquet-free route to momentum-space selection of temporal stability, enabling polarization- and direction-selective control over oscillatory, frozen, and exponentially growing/decaying states.

## 2.Results:

### 2.1 Scattering Dynamics at a Single Temporal Boundary in Hyperbolic

## Metamaterial

We begin by comparing the conservation laws at spatial and temporal boundaries. We consider nonmagnetic media ($\mu_r = 1$) with strictly real-valued permittivity. In Fig. 1A, vacuum ($\varepsilon_1 = 1$) is adjacent to a hyperbolic medium $\varepsilon_2 = \mathrm{diag}(\varepsilon_x, \varepsilon_y, \varepsilon_z)$, with $\varepsilon_x < 0$ and $\varepsilon_y, \varepsilon_z > 0$. At the spatial boundary, the frequency and tangential wavevector component are conserved, so the hyperbolic equifrequency contours (EFCs) determine the matched refracted state and allows negative refraction through wavevector matching [28–30]. The temporal counterpart obeys different conservation laws. For a spatially uniform temporal modulation, the total wavevector $k$ is conserved, whereas the breaking of temporal translation symmetry allows for frequency changes. Material dispersion particularly significant for passive hyperbolic media [31–35]. To isolate the effect of the indefinite permittivity tensor itself, we therefore adopt a hyperbolic response that is approximately frequency-independent. This idealization should be understood as an actively sustained effective response rather than a passive hyperbolic medium.
We consider a spatially uniform and unbounded medium that switches abruptly at $t = t_1$ from vacuum ($\varepsilon_1 = 1$) to the hyperbolic tensor $\varepsilon_2 = \mathrm{diag}(\varepsilon_x, \varepsilon_y, \varepsilon_z)$ in Fig. 1B. The switching time is assumed to be much shorter than the incident optical period, so that the change can be treated as an ideal temporal boundary. Before introducing a specific incident field, the behavior of the electromagnetic field beyond the boundary can be classified directly based on Maxwell's equations.
For a p-polarized spatial harmonic wave, the magnetic field may be written as

$$H_z(\boldsymbol{r}, t) = H_{\boldsymbol{k}}(t) e^{i\boldsymbol{k}\cdot\boldsymbol{r}} \tag{1}$$

where $\boldsymbol{k} = (k_x, k_y)$ is conserved across the temporal boundary. Substitution into Maxwell's equations gives

$$\frac{d^2 H_{\mathbf{k}}}{dt^2} + \Omega_p^2(\boldsymbol{k}) H_{\boldsymbol{k}} = 0 \tag{2}$$

with

$$\Omega_p^2(\boldsymbol{k}) = c^2 \left(\frac{k_x^2}{\varepsilon_y} + \frac{k_y^2}{\varepsilon_x}\right) = c^2 k^2 \left(\frac{\cos^2\theta}{\varepsilon_y} + \frac{\sin^2\theta}{\varepsilon_x}\right). \tag{3}$$

Equation (2) has the mathematical form of a temporal oscillator and $\Omega_p^2$ acts as an effective temporal stiffness. The post-boundary temporal eigenfrequency $\omega_{2p}$ is defined through $\omega_{2p}^2(\boldsymbol{k}) = \Omega_p^2(\boldsymbol{k})$. Because $\boldsymbol{k}$, $\varepsilon_x$ and $\varepsilon_y$ are real in the present model, $\Omega_p^2$ is strictly real. Consequently, the temporal eigenfrequency $\omega_{2p}$ beyond the temporal boundary can only be real, zero, or purely imaginary. For the hyperbolic medium considered here, the sign of $\Omega_p^2$ varies only with the wavevector direction $\theta$. Taking the conventional case first, $\Omega_p^2 > 0$ gives $\omega_{2p} = \pm\Omega_p$ and conventional temporal refraction and reflection occur. At the critical direction $\Omega_p^2(\theta_0) = 0$ and $d^2 H_{\boldsymbol{k}}/dt^2 = 0$, combining the temporal boundary condition, the mode increasing linearly with time is absent and magnetic field distribution is time-independent. Finally, when $\Omega_p^2 < 0$, we define $\gamma = \sqrt{-\Omega_p^2}$, so that $\omega_{2p} = \pm i\gamma$. The two temporal modes vary as $\exp[\pm\gamma(t - t_1)]$. The growing mode produces exponential amplification while the spatial phase factor $\exp(i\boldsymbol{k}\cdot\boldsymbol{r})$ remains fixed. Fig. 1B can therefore be regarded as an angular phase map of this effective temporal stiffness, separating the oscillatory, critical and exponentially evolving regimes.
Within the same temporal oscillator framework, the polarization selectivity follows directly from the corresponding stiffness. For s polarization, the corresponding effective temporal

stiffness is

$$\Omega_s^2 = \frac{c^2k^2}{\varepsilon_z} \tag{4}$$

With $\varepsilon_z > 0$, $\Omega_s^2$ remains positive for every propagation direction, so the s-polarized mode stays in the real-frequency regime and exhibits only conventional temporal scattering. Therefore, the contrast with p-polarization stems directly from the unique anisotropy of the in-plane tensor, complementing earlier anisotropic temporal effects such as the temporal Brewster response, temporal aiming, and polarization conversion [10,14,15].

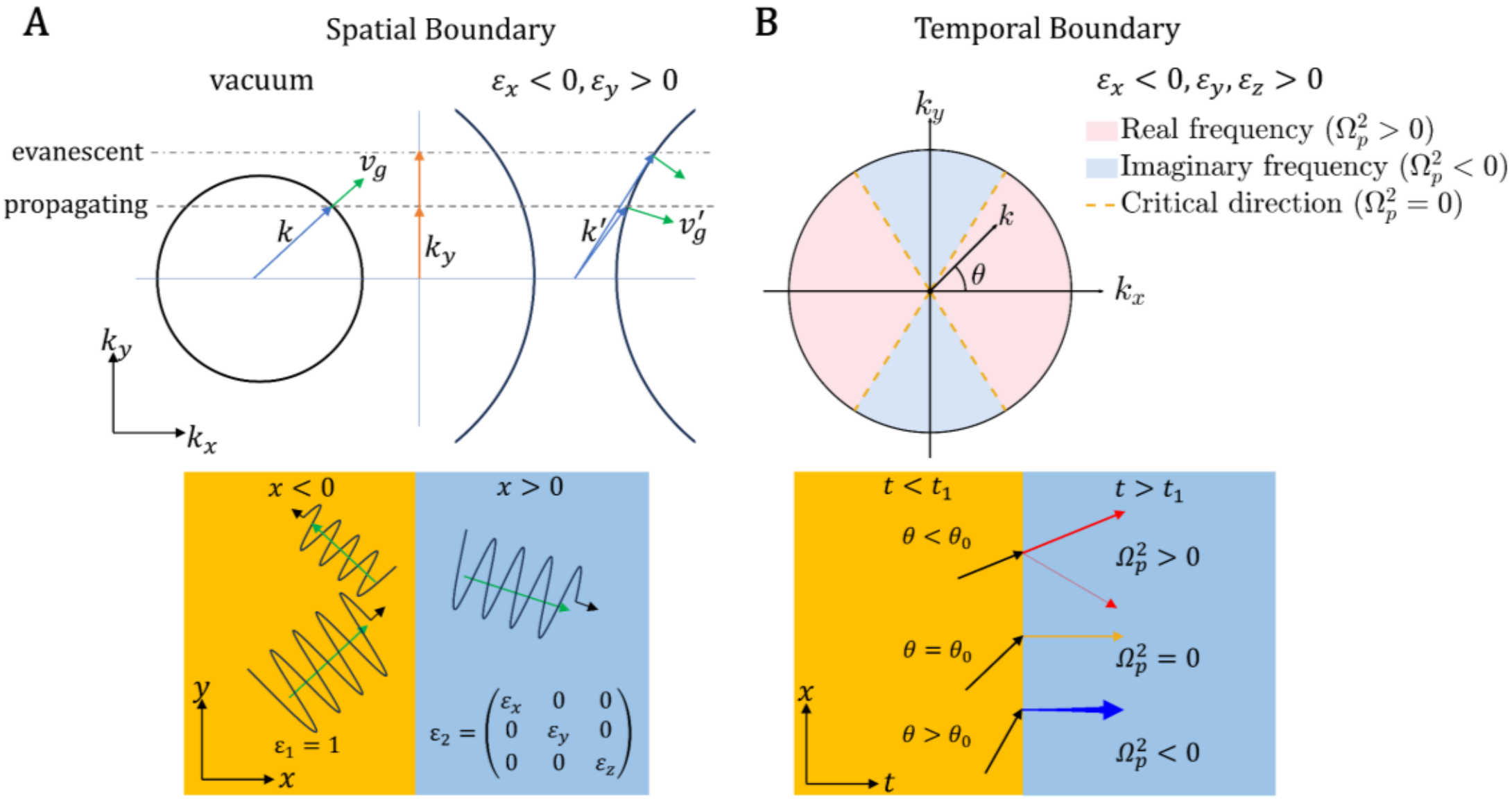

**Figure 1**. Conservation laws and modal classification at spatial and temporal boundaries involving the same hyperbolic medium. (A) Equifrequency contours (EFCs) of a spatial boundary at $x = 0$ between vacuum ($\varepsilon_1 = 1$) and a hyperbolic medium $\varepsilon_2 = \mathrm{diag}(\varepsilon_x, \varepsilon_y, \varepsilon_z)$, with $\varepsilon_x < 0$ and $\varepsilon_y > 0$. The conservation of frequency and the tangential wavevector component $k_y$ determines the matched states on hyperbolic EFC. Besides a propagating incident component ($|k_y| < k_0$), a high-$k_y$ component that is evanescent in vacuum can also intersect the open hyperbolic EFC and thereby couple to a propagating bulk mode. Green arrows denote the group velocities normal to the EFCs. The figure below illustrates the negative refraction of p-polarized waves. (B) Temporal boundary at $t = t_1$ for the corresponding vacuum-to-hyperbolic switch. Spatial homogeneity conserves the total wavevector $k$. The upper angular phase map is determined by the sign of $\Omega_p^2$: red sectors, $\Omega_p^2 > 0$ (real frequency regime); yellow dashed boundaries, $\Omega_p^2(\theta_0) = 0$ (zero frequency regime); bule sectors, $\Omega_p^2 < 0$ (imaginary frequency regime). The figure below shows the temporal evolution of the corresponding p-polarization waves with different wavevector directions.

To relate this modal classification to the field generated by a temporal switch, we consider a monochromatic p-polarized plane wave as a specific example of boundary matching. Before the boundary, the magnetic field is $H_{z1} = e^{-i\omega_1 t + ik_x x + ik_y y}$. Here $\omega_1$ is the incident angular frequency and $k_x, k_y$ are the $x$- and $y$- components of the conserved wavevector. For $t > t_1$, the field is the superposition of two temporal modes. In the real frequency regime these are conventionally identified as the time-refracted and time-reflected waves; in the imaginary frequency regime the pair of modes is extended to the exponential decreasing and the exponential increasing branches. Assuming their amplitudes as $T_1$ and $R_1$ respectively, gives

$$H_{z2} = T_1 e^{-i\omega_2(t-t_1)-i\omega_1 t_1+ik_x x+ik_y y} + R_1 e^{i\omega_2(t-t_1)-i\omega_1 t_1+ik_x x+ik_y y} \tag{5}$$

The modal frequency $\omega_2$ is fixed by the hyperbolic dispersion, whereas the amplitudes are fixed by the temporal boundary conditions. Continuity of the electric displacement **D** and magnetic flux intensity **B** across the temporal boundary, alongside the conservation of wavevector [1,2], yields:

$$T_1 = 1/2\,(1 + \omega_2/\omega_1) \tag{6.1}$$

$$R_1 = 1/2\,(1 - \omega_2/\omega_1) \tag{6.2}$$

with $\omega_1 = k_0 c/\sqrt{\varepsilon_1}$ and $\omega_2 = c\sqrt{k_x^2/\varepsilon_y + k_y^2/\varepsilon_x} = ck_0\sqrt{\cos(\theta)^2/\varepsilon_y + \sin(\theta)^2/\varepsilon_x}$. The conserved wavevector components are $k_x = k_0\cos(\theta), k_y = k_0\sin(\theta)$, with $\theta$ measured from the $x$-axis. Therefore $\Omega_p^2 = \omega_2^2$, and the plane-wave solution is a direct representation of the angular phase map. In the real frequency region, the two modes remain oscillatory; at the critical direction their sum yields the critical magnetic-field freezing determined by the temporal boundary conditions; and for a purely imaginary $\omega_2 = i\gamma$, they become exponentially decaying and growing temporal modes.

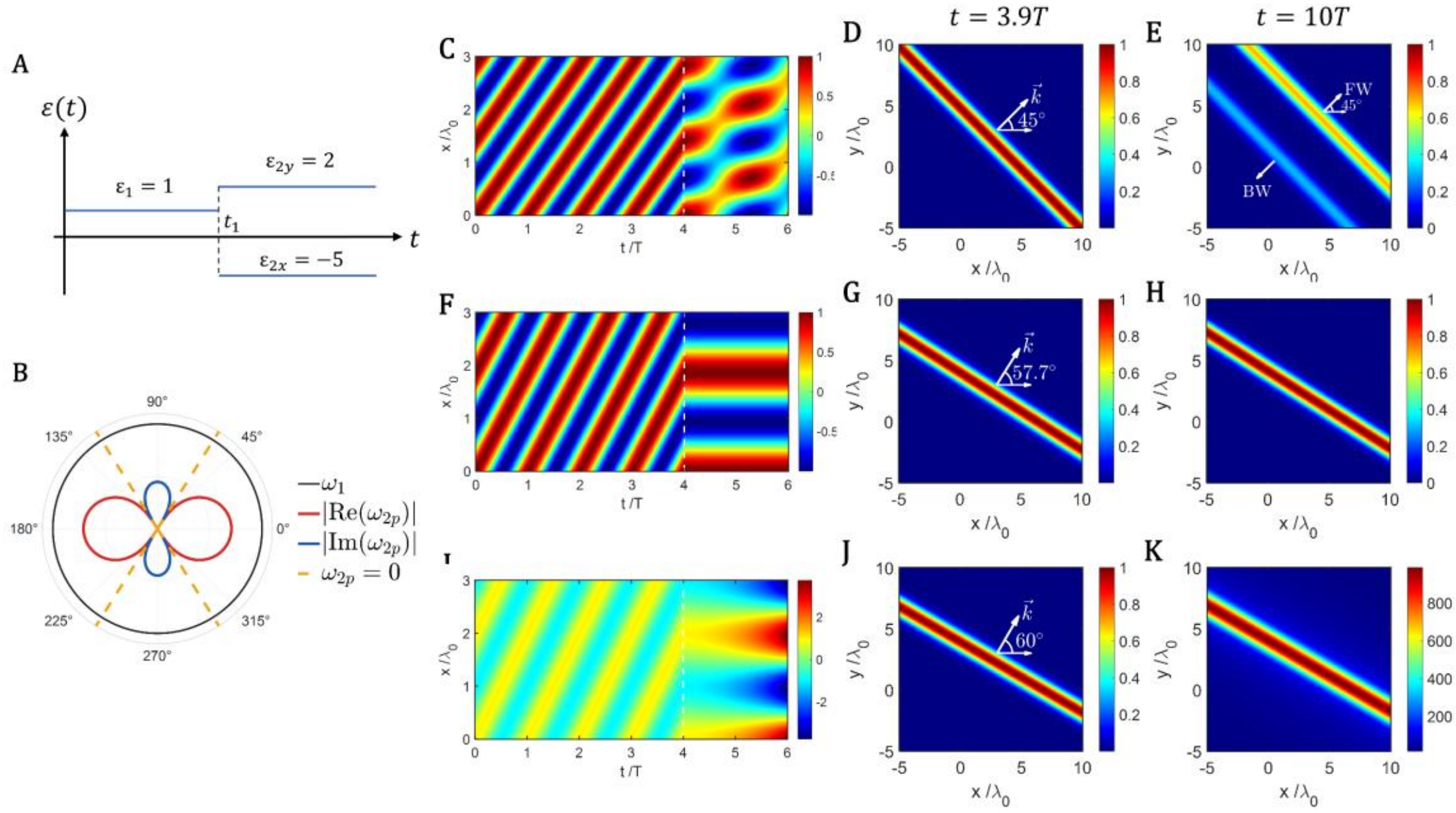

**Figure 2**. Direction-dependent temporal scattering of p-polarized waves at a single hyperbolic temporal boundary. (A) Step-like temporal modulation of the permittivity at $t_1 = 4T$ ($T = \lambda_0/c$, $\lambda_0 = 1um$, $\omega_1 = 2\pi c/\lambda_0 \approx 1.885\times 10^{15} rad/s$), where the medium switches from vacuum $\varepsilon_1 = 1$, to a hyperbolic state with $\varepsilon_2 = \text{diag}(-5,2,2)$. And the critical angle is $\theta_0 = tan^{-1}(\sqrt{-\frac{\varepsilon_x}{\varepsilon_y}}) \approx 57.7°$. (B) Equi-wavevector contours (EWCs) showing the angular dependence of the converted frequency for a conserved wavevector. The black circle denotes the normalized incident frequency $\omega_1$; the red and blue curves represent $|\text{Re}(\omega_{2p})|$ and $|\text{Im}(\omega_{2p})|$, respectively, while the yellow-dashed lines indicate the zero-frequency critical directions $\omega_{2p} = 0$. These contours separate the wavevector directions into real-frequency, zero-frequency and purely imaginary-frequency regimes. (C-E) Real-frequency regime at $\theta = 45° < \theta_0$ ($\omega_{2p} \approx 7.300\times 10^{14} rad/s$): the plane wave spatiotemporal evolution (C) exhibits conventional oscillatory temporal scattering, while Gaussian pulse snapshots immediately before ($t = 3.9T$) and after ($t = 10T$) the switch show the resulting forward- (FW) and backward-propagating (BW)

components. (F-H) Zero-frequency critical regime at $\theta = \theta_0$: after the temporal boundary, the plane wave becomes time independent (F), and the Gaussian pulse remains spatially localized (G, H), demonstrating magnetic field freezing. (I-K) Purely imaginary-frequency regime at $\theta = 60° > \theta_0$ ($\omega_2 \approx 2.980 \times 10^{14} i\ rad/s$): the post-boundary plane wave ceases to propagate and evolves exponentially in time (I), while the Gaussian pulse snapshots (J, K) show negligible spatial displacement accompanied by strong field amplification.

As illustrated in Fig. 2A, we assume a rapid, uniform transition of the unbounded medium from vacuum ($\varepsilon_1 = 1$) to an anisotropic hyperbolic material at $t_1 = 4T$. Fig. 2B provides the Equi-wavevector contours (EWCs) representation of the same modal classification summarized by the angular phase map in Fig. 1B. Accordingly, the red, yellow-dashed, and blue regions in Fig. 2B correspond directly to $\Omega_p^2 > 0$, $\Omega_p^2 = 0$, and $\Omega_p^2 < 0$, respectively. At the temporal boundary, the wavevector is conserved; therefore, changing the angle $\theta$ alters the converted frequency $\omega_2(\theta)$. Unlike Fig. 1B which shows only the sign of the effective stiffness, the radial coordinates in Fig. 2B represent frequency amplitudes. For $\varepsilon_2 = \mathrm{diag}(-5,2,2)$, the critical direction $\theta_0$ satisfies $\Omega_p^2(\theta_0) = 0$ ($\theta_0 \approx 57.7°$). The red EWC curve corresponds to the real frequency regime, the curve collapses to zero at the yellow dashed critical directions, and the blue curve plots $|Im(\omega_2)|$ in the purely imaginary regime. This quantitative EWCs representation motivates the three representative directions used below.

To connect the single-harmonic solution with a localized excitation, we evaluate both a continuous plane wave and a narrowband Gaussian pulse. The pulse is decomposed into spatial harmonics in k space. Because each k component is conserved across the temporal boundary, the temporal matching conditions are applied independently to every component, after which the total field is reconstructed by an inverse spatial Fourier transform.

We first consider the conventional real-frequency case, $\theta = 45° < \theta_0$, for which $\Omega_p^2 > 0$. Figs. 2C–2E shows that the field beyond the boundary exhibits an oscillatory temporal phase evolution, with standard temporal refraction and reflection occurring. The conservation of wavevectors is accompanied only by frequency conversion; neither a critical magnetic-field freezing nor exponential amplification occurs. This real-frequency response provides a reference for identifying critical and unstable regions.

At the critical direction $\theta = \theta_0$, $\Omega_p^2$ vanishes and hence $\omega_2 = 0$. The results for plane waves and Gaussian pulses shown in Figs. 2F–2H indicate that the magnetic field distribution becomes a steady state beyond the temporal boundary: it retains its spatial periodicity but no longer translates in space. Although $\boldsymbol{H}$ remains constant over time, its spatial curl is still nonzero ($\nabla \times \boldsymbol{H} \neq 0$). Therefore, as required by Maxwell's equations, the $\partial \boldsymbol{D}/\partial t$ is a nonzero constant. Thus, for $t > t_1$ both **E** and **D** increase linearly with time. The detailed derivation is provided in the Supplementary Information. This zero-frequency state represents the critical boundary between oscillatory and exponential dynamics.

Finally, for $\theta = 60° > \theta_0$, $\Omega_p^2 < 0$ and the converted frequency is purely imaginary. Figs. 2I–2K show that the oscillatory temporal phase is replaced by growing and decaying exponential modes. Because the group velocity is zero, the field pattern does not undergo spatial translation after the switch, while the growing branch increasingly dominates the total field. The Gaussian pulse snapshots show negligible displacement together with a strong increase in field amplitude, directly demonstrating the direction-selective temporal evanescence and amplification predicted by the effective temporal oscillator model.

## 2.2 Dielectric-Hyperbolic-Dielectric Temporal Slab Dynamics

We next extend the single-boundary dynamics to a finite duration dielectric-hyperbolic-dielectric temporal slab. As illustrated in Fig. 3A, the initial isotropic dielectric medium ($\varepsilon_1 = 1$) is abruptly switched at $t_1 = 4T$ to the hyperbolic state with $\varepsilon_2 = diag(-5,2,2)$, and is restored to the initial dielectric state $\varepsilon_3 = \varepsilon_1 = 1$ at $t_2 = 6.9T$. The interval ($\tau = t_2 - t_1$) defines the temporal thickness of the hyperbolic slab. Such a double-temporal-boundary configuration is the temporal counterpart of a finite spatial layer, for which the duration $\tau$ controls the evolution inside the slab [36,37]. Since the medium after the second temporal boundary is identical to the initial medium, conservation of the wavevector ensures that the waves released from the slab return to the initial dispersion and hence to the original frequency. The role of the second boundary is to convert the state bounded inside the hyperbolic temporal slab into propagating modes.

The three angular regimes identified at the single-boundary can be described within the transfer framework. For each p-polarized spatial harmonic, the magnetic inside the hyperbolic slab obeys Eq. (2). Introducing the state vector $\mathbf{X}(t) = \begin{pmatrix} H_{\mathbf{k}}(t) \\ \dot{H}_{\mathbf{k}}(t) \end{pmatrix}$, its evolution between the two boundaries can be written as

$$\mathbf{X}(t_2) = \mathbf{P}(\Omega_p, \tau)\, \mathbf{X}(t_1), \tag{7}$$

where

$$\mathbf{P}(\Omega_p, \tau) = \begin{cases} \begin{pmatrix} \cos(\Omega_p \tau) & \dfrac{\sin(\Omega_p \tau)}{\Omega_p} \\ -\Omega_p \sin(\Omega_p \tau) & \cos(\Omega_p \tau) \end{pmatrix}, & \Omega_p^2 > 0, \\ \begin{pmatrix} 1 & \tau \\ 0 & 1 \end{pmatrix}, & \Omega_p = 0, \\ \begin{pmatrix} \cosh(\gamma\tau) & \dfrac{\sinh(\gamma\tau)}{\gamma} \\ \gamma \sinh(\gamma\tau) & \cosh(\gamma\tau) \end{pmatrix}, & \Omega_p = i\gamma, \quad \gamma > 0. \end{cases} \tag{8}$$

Here, $\mathbf{P}$ describes the evolution within the hyperbolic slab, whereas the temporal boundary conditions at $t_1$ and $t_2$ determine how the modes of dielectric are coupled to and from the internal state in hyperbolic medium. Therefore, the three different behaviors shown in Fig. 3 are specific manifestations of distinct evolutionary processes.

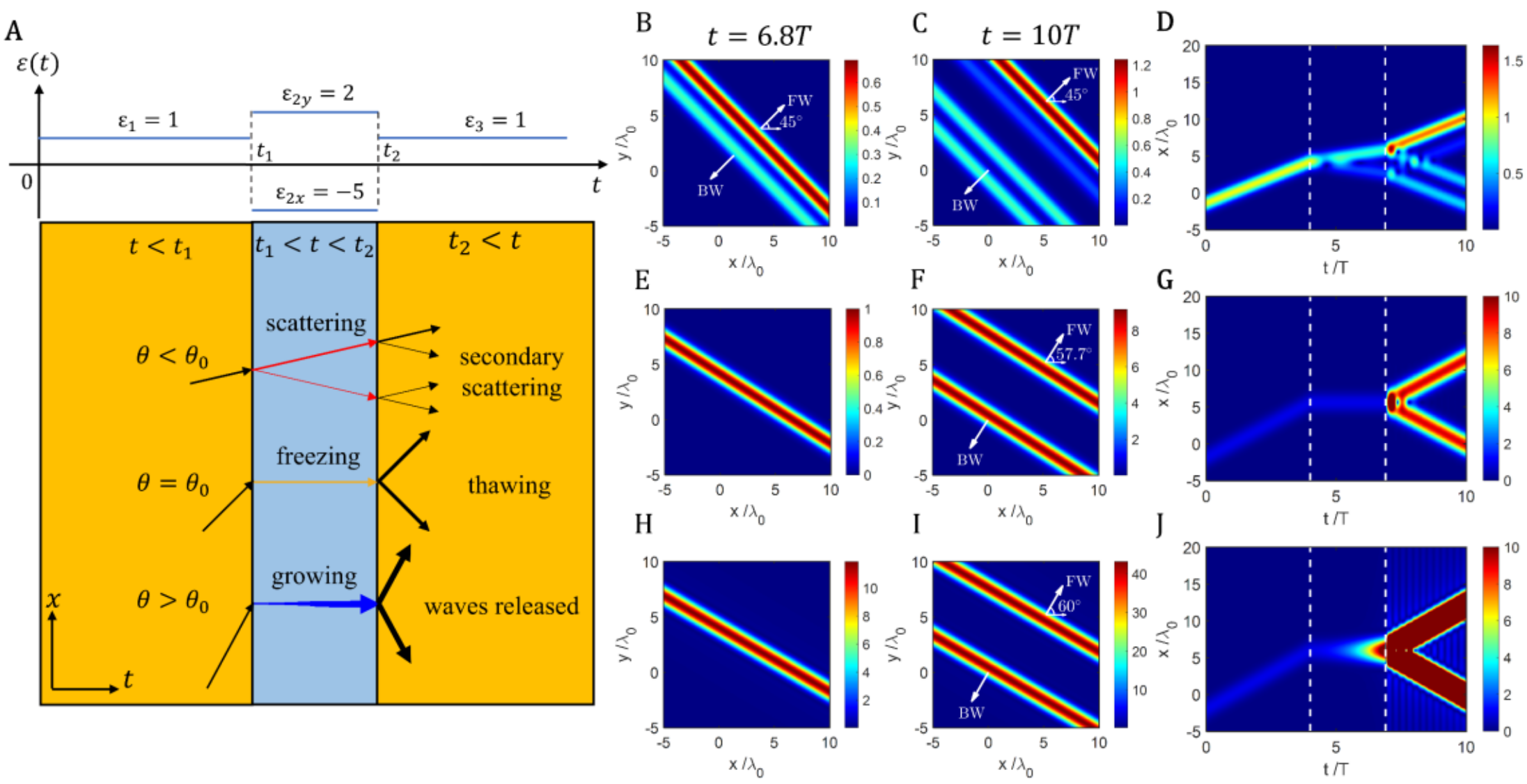

**Figure 3**. Spatiotemporal evolution of p-polarized pulses across a dielectric-hyperbolic-dielectric temporal slab. (A) Permittivity profile and schematic wave dynamics. The medium switches from the initial dielectric $\varepsilon_1 = 1$ to a hyperbolic state $\varepsilon_2 = diag(-5,2,2)$ at $t_1 = 4T$, and returns to $\varepsilon_3 = 1$ at $t_2 = 6.9T$; the slab duration is $\tau = t_2 - t_1$, and the corresponding critical angle is $\theta_0 \approx 57.7°$. (B-D) Real frequency regime ($\theta = 45° < \theta_0, \Omega_p^2 > 0$): instantaneous magnetic-field distributions inside the slab (B) and after the second temporal boundary (C), together with the full spatiotemporal evolution (D), showing conventional temporal scattering between the two temporal boundaries. (E-G) Zero frequency critical direction ($\theta = \theta_0, \Omega_p^2 = 0$): the magnetic field distribution remains spatially frozen inside the slab (E) and is released into forward- and backward-propagating waves when the initial dielectric is restored (F); the corresponding freezing-and-thawing process is shown in (G). (H-J) Purely imaginary-frequency regime ($\theta = 60° > \theta_0$, $\Omega_p^2 < 0$): the spatially non-propagating field undergoes exponential amplification inside the slab (H) and is subsequently released as strongly amplified forward- and backward-propagating waves (I), as visualized in the complete spatiotemporal evolution (J). White dashed lines in (D), (G), and (J) indicate the two temporal boundaries. FW and BW denote the forward- and backward-propagating components, respectively.

For $\theta < \theta_0$, $\Omega_p^2 > 0$ and the evolution inside the slab remains propagation. The first temporal boundary generates the conventional time-refracted and time-reflected waves shown in Fig. 3B. During the slab, these two components propagate inversely and the second boundary transfer them into two forward- and two backward-propagating modes as shown in Fig. 3C. Consequently, the output amplitudes depend on the accumulated phase $\Omega_p \tau$, producing the temporal analogue of Fabry-Pérot interference in a temporal slab [36,37]. The complete evolution in Fig. 3D shows the two temporal scattering and the four output propagating modes. It is important to understand multiple scattering paths as coherent contributions to the final forward- and backward-modes, rather than as four independent output beams.

As the propagation direction equal to the critical angle $\theta = \theta_0$, the effective stiffness vanishes, and the propagator reduces to zero-frequency regime. For the temporal boundary conditions considered here, the linear-in-time magnetic mode is absent, leaving a temporally constant magnetic field, as shown in Fig. 3E. Importantly, magnetic-filed freezing does not imply that the entire electromagnetic state is static. Because $H_z$ retains its spatial dependence, $\nabla \times \boldsymbol{H} \neq 0$, and $\partial \boldsymbol{D}/\partial t = \nabla \times \boldsymbol{H}$. Consequently, the transverse electric displacement accumulates linearly during the slab. At $t = t_2$, restoring the initial dielectric couples the frozen (magnetic field) state to propagating modes. The field is consequently thawed into forward- and backward- propagating waves at the original frequency, as shown in Fig. 3F, while the spatiotemporal dynamics of Fig. 3G directly visualizes the frozen interval followed by wave recovery. The "thawing" is more precisely interpreted as the release of a critical zero frequency state by the second temporal boundary rather than to the recovery of an entirely static electromagnetic field.

For $\theta > \theta_0$, $\Omega_p^2 < 0$ and $\Omega_p = i\gamma$, the trigonometric evolution of the propagating regime is transferred into the hyperbolic function $\cosh(\gamma\tau)$ and $\sinh(\gamma\tau)$. The internal temporal modes are therefore exponentially decaying and growing rather than oscillatory. Due to the imaginary frequency, the field remains spatially localized while its amplitude increases as the growing branch progressively dominates, as illustrated in Fig. 3H. At the second temporal boundary, this amplified non-propagating state is converted back into two propagating modes supported by the dielectric, producing the strongly enhanced forward and backward waves in Fig. 3I and the complete evolution as shown in Fig. 3J. In this regime, the temporal thickness controls an exponential transfer factor $e^{\gamma\tau}$ and $e^{-\gamma\tau}$. The finite interval of imaginary frequency is formally similar to the transmission through a limited spatial layer; however, unlike passive spatial tunneling, this actively sustained temporal medium allows for the emergence of a continuously growing mode, thereby enabling enhanced transmission rather than merely decaying transmission.

The dielectric–hyperbolic–dielectric temporal slab thus converts the direction-dependent evolution established at a single temporal boundary into a finite-duration transfer process. As $\Omega_p^2$ changes from positive to zero and then negative, the internal evolution changes from oscillatory temporal interference, through critical freezing and thawing, to exponential amplification and release. The second boundary is not merely an additional scattering, but as the mechanism that converts the frozen or amplified states generated inside the hyperbolic interval back into propagating waves.

## 2.3 FDTD Numerical Simulation Verification

To provide an independent numerical validation of the analysis in Section 2.1 and 2.2, we performed finite-difference time-domain (FDTD) numerical simulations of a narrowband Gaussian pulse propagating along the $x$-axis in a spatially uniform dielectric-hyperbolic-dielectric temporal slab (the direction is chosen such that the p-polarized wave lies in the purely imaginary-frequency regime, whereas the s-polarized pulse remains in the real-frequency regime). The medium switches from vacuum ($\varepsilon_1 = 1$) to a hyperbolic material $\varepsilon_2 = diag(2, -5, 2)$ at time $t_1 = 20\mathrm{T}$ ($T = \frac{\lambda_0}{c} = 3.336 fs$), and returns to vacuum at $t_2 = t_1 + \tau$.

We compare two temporal slab durations $\tau_1$ and $\tau_2$ to validate the duration dependence of the p-polarized amplification and the polarization selectivity predicted by the analytical model.

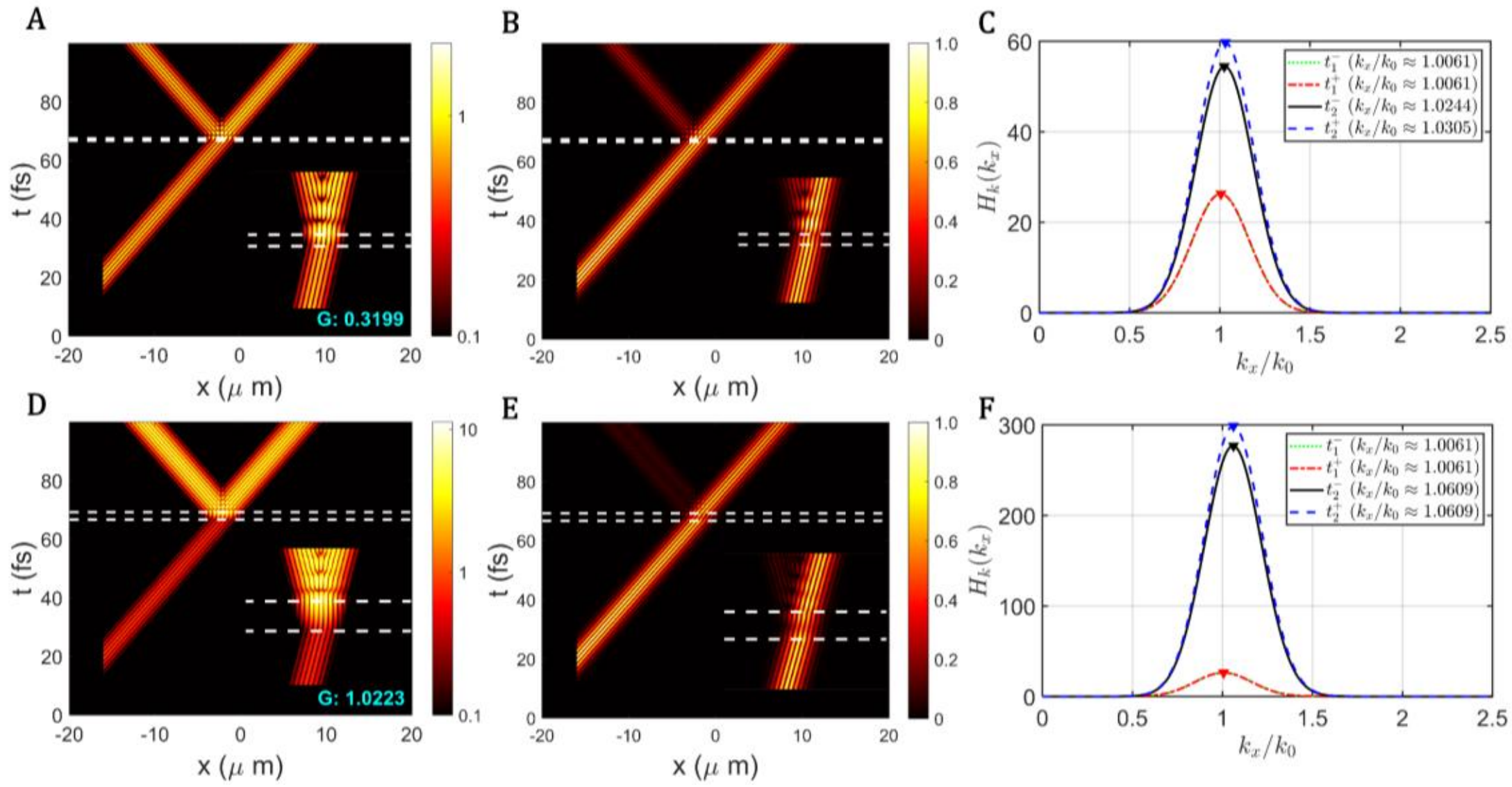

**Figure 4**. FDTD validation of polarization-selective dynamics in a hyperbolic temporal slab. Spatiotemporal and k-space evolutions are shown for two slab durations, $\tau_1 = 0.850fs$ (A–C) and $\tau_2 = 2.550fs$ (D–F), with white dashed lines marking the two temporal boundaries. (A, D) Spatiotemporal maps of the p-polarized Gaussian pulse $|H_z|$ for a direction satisfying $\Omega_p^2 < 0$. The pulse becomes spatially non-propagating inside the hyperbolic interval while its amplitude grows exponentially; $G(\tau) = log_{10}(\max|H_z(x, t_2^-)|/\max|H_z(x, t_1^+)|)$. (B, E) Corresponding s-polarized results $|E_z|$ for the same direction, for which $\Omega_s^2 > 0$ and the pulse remains in the oscillatory real-frequency regime. (C, F) k-space spectra of the p-polarized pulse near the two temporal boundaries $t_1$ and $t_2$, showing the overall amplification and the redistribution of spectral weight caused by k-dependent temporal amplification. The position of the dominant spectral peak is shown in the legend.

Figures 4A and 4D show the spatiotemporal evolution of the p-polarized pulse. Immediately after the first temporal boundary, the pulse remains centered at nearly the same spatial position while its field amplitude increases strongly with time. This behavior is the direct numerical signature of the purely imaginary-frequency regime: $\Omega_p = i\gamma$, and the two temporal modes vary as $\exp[\pm\gamma(t - t_1)]$. The observed state should therefore be distinguished from the zero-frequency critical state discussed in Section 2.1: here the field pattern is spatially non-propagating, but its amplitude is not static and instead becomes dominated by the growing mode. For the central Fourier component $k_x = k_0$, the growth gives $\gamma \approx k_0 c/\sqrt{5}$. This corresponds to ideal amplitude factors of approximately 2.05 and 8.58 for $\tau_1$ and $\tau_2$, respectively. To quantify this growth, the plotted quantity should be defined as a normalized logarithmic gain, and a convention definition is $G(\tau) = log_{10}(\max|H_z(x, t_2^-)|/\max|H_z(x, t_1^+)|)$. In the exponential regime, $G(\tau) \approx \frac{\gamma\tau}{\ln 10}$, so that $\frac{G(\tau_2)}{G(\tau_1)} \approx 3$ for ideal results. For the FDTD simulation, the corresponding values of $Gain$ are approximately 0.3199 and 1.0223. At the second temporal boundary, the amplified state is converted back into propagating modes of the final dielectric, producing forward- and backward-propagating wave packets. For a sufficiently long slab duration, the growing mode dominates before the second switch, consistent with the nearly symmetric release of the amplified field shown in Fig. 3J.

The s-polarized response provides a contrast case as shown in Figs. 4B and 4E. The pulse continues to translate spatially within the hyperbolic interval and does not exhibit the exponential increase observed for p-polarization. This case follows from the polarization-dependent temporal stiffness derived in Section 2.1: for the wavevector direction, $\Omega_s^2 > 0$, the s-polarized temporal modes remain oscillatory rather than growing and decaying. The second temporal boundary subsequently converts the field back into propagating content in the final dielectric. Multiple spatially separated trajectories can be observed after two temporal scattering events, and it is because these trajectories arise from multiple wave packet branches generated by coherent temporal scattering paths.

Figs. 4C and 4F show the k-space spectra of the p-polarized Gaussian pulses at the moments near the two temporal boundaries for two slab durations $\tau_1$ and $\tau_2$. Since the time modulation is spatially uniform, each spatial Fourier component maintains its wavevector at the two temporal boundaries. Therefore, the shift of the dominant spectral peak towards a larger $k_x$ should not be interpreted as the change of the wavevector of an individual Fourier component at the temporal boundaries, nor should it be regarded as a frequency blue shift. Instead, the finite bandwidth incident pulse contains a distribution of conserved k-components whose temporal growth rates are generally different. In the growing regime, the dominant contribution evolves approximately as

$$H_k(k, t > t_1) \propto H_k(k, t_1) e^{\gamma(k)(t-t_1)} \tag{9}$$

where $\gamma(k) = \frac{kc}{\sqrt{|\varepsilon_y|}}$ [23]. The k-dependent gain consequently reweights the spectra, so that the dominant spectral weight can move toward larger wavevector even though every individual k-component remains wavevector-conserved (in the Supplementary Information). A longer time interval will enhance this spectral reweighting effect and also bring about an overall amplification. The second temporal boundary mainly converts the amplified internal state back into propagating modes, while the spatial wavevector remains conserved across the boundary.

## 3.Conclusion

In summary, we have established a polarization- and direction-selective temporal paradigm enabled by a single dielectric-hyperbolic temporal boundary. For p-polarized waves, varying only the wavevector direction drives a transition from conventional oscillatory temporal scattering, through a zero-frequency critical state with a stationary magnetic field pattern, to a purely imaginary frequency regime characterized by spatially non-propagating fields with exponentially growing and decaying temporal modes. In contrast, the s-polarized waves remain in the real-frequency regime for all directions, demonstrating that the evolution is intrinsically both polarization and direction selective.

Extending the system to a dielectric-hyperbolic-dielectric temporal slab further show that a second temporal boundary converts the zero- or imaginary frequency states back into propagating waves in the final dielectric. The analytical temporal boundary model, spatial-harmonic reconstruction of Gaussian pulses, and FDTD simulations consistently verify these polarization- and direction-dependent dynamics, and reveal a temporal counterpart of the spatially evanescent-propagating transition associated with hyperbolic dispersion. Hyperbolic temporal boundaries therefore provide a compact framework for controlling the temporal stability, amplification, and release of electromagnetic fields in momentum space, opening opportunities for polarization- and direction-selective wave manipulation in time-varying metamaterials.

## Acknowledgments

This work was supported by National Natural Science Foundation of China (Grant Nos. 12404371 and 12361161667), Shenzhen Science and Technology Program (JCYJ20250604122934046), Fujian Provincial Natural Science Foundation of China (2024J011002). Zhao Wang acknowledges the support from the China Scholarship Council (202406310168)

# 1. Temporal Transmission and Reflection Coefficients for Non-Critical Incidence

## 1.1 Scattering at the First Temporal Boundary

The first temporal boundary is induced at $t = t_1$, where the unbounded medium's permittivity abruptly transitions from vacuum ($\varepsilon_1 = 1$) to a hyperbolic medium characterized by the dielectric tensor $\varepsilon_2 = diag(\varepsilon_x, \varepsilon_y, \varepsilon_z)$ . Provided this transition occurs on a timescale significantly shorter than the oscillation period of the incident waves, thus it constitutes an ideal temporal boundary. We consider a p-polarized incident plane wave described by the following magnetic and electric displacement fields:

$$H_{z1} = e^{-i\omega_1 t + ik_x x + ik_y y} \tag{1}$$

$$D_{x1} = -\frac{k_y}{\omega_1} e^{-i\omega_1 t + ik_x x + ik_y y} \tag{2}$$

Here $\omega_1$ is the initial angular frequency, and $k_x, k_y$ are the wavevector components along the $x$- and $y$- axes, respectively. Let $T_1$ and $R_1$ denote the temporal transmission and reflection coefficients. For $t > t_1$, the total magnetic field is the superposition of the two components:

$$H_{z2} = T_1 e^{-i\omega_2 (t-t_1) - i\omega_1 t_1 + ik_x x + ik_y y} + R_1 e^{i\omega_2 (t-t_1) - i\omega_1 t_1 + ik_x x + ik_y y} \tag{3}$$

The corresponding electric displacement field D along the $x$-axis is derived as:

$$D_{x2} = \frac{k_y}{\omega_2} e^{-i\omega_1 t_1 + ik_x x + ik_y y} \left(-T_1 e^{-i\omega_2 (t-t_1)} + R_1 e^{i\omega_2 (t-t_1)}\right) \tag{4}$$

Because spatial translation symmetry is strictly maintained, the wavevector is conserved across the temporal boundary, while $\omega_2$ represents the converted frequency in the hyperbolic medium. The fundamental boundary conditions require the continuity of the electric displacement field **D** and magnetic flux intensity **B** across the temporal boundary at $t = t_1$:

$$B_{z1}(t_1^-) = B_{z2}(t_1^+) \tag{5.1}$$

$$D_{x1}(t_1^-) = D_{x2}(t_1^+) \tag{5.2}$$

Assuming non-magnetic materials throughout the process, the continuity of **B** directly enforces the continuity of the magnetic field H. Substituting the field expressions evaluated at the boundary into these continuity equations yields the following:

$$1 = T_1 + R_1 \tag{6.1}$$

$$-\frac{1}{\omega_1} = \frac{1}{\omega_2}(-T_1 + R_1) \tag{6.2}$$

Solving this system analytically provides the explicit expressions for the temporal transmission and reflection coefficients at the first boundary:

$$T_1 = 1/2\,(1 + \omega_2/\omega_1) \tag{7.1}$$

$$R_1 = 1/2\,(1 - \omega_2/\omega_1) \tag{7.2}$$

### 1.2 Scattering at the Second Temporal Boundary

The second boundary is induced at $t = t_2$, where the medium's permittivity abruptly reverts from the hyperbolic state back to the initial value $\varepsilon_3 = 1$. Let $T_2$ and $R_2$ denote the temporal transmission and reflection coefficients at this second interface. Upon encountering this boundary, the existing time-refracted and time-reflected waves split, generating a total of two forward-propagating and two backward-propagating beams. For $t > t_2$ the total magnetic field is expressed as:

$$H_{z3} = \left[\left(T_1 T_2 e^{-i\omega_2\tau} + R_1 R_2 e^{i\omega_2\tau}\right)e^{-i\omega_1(t-t_2)}\right.$$

$$\left.+\left(T_1 R_2 e^{-i\omega_2\tau} + R_1 T_2 e^{i\omega_2\tau}\right)e^{i\omega_1(t-t_2)}\right]e^{-i\omega_1 t_1 + ik_x x + ik_y y} \qquad (8)$$

The corresponding electric displacement field D along the $x$-axis for $t > t_2$ is given by:

$$D_{x3} = \frac{k_y}{\omega_1}\left[-\left(T_1 T_2 e^{-i\omega_2\tau} + R_1 R_2 e^{i\omega_2\tau}\right)e^{-i\omega_1(t-t_2)}\right.$$

$$\left.+\left(T_1 R_2 e^{-i\omega_2\tau} + R_1 T_2 e^{i\omega_2\tau}\right)e^{i\omega_1(t-t_2)}\right]e^{-i\omega_1 t_1 + ik_x x + ik_y y} \qquad (9)$$

where $\tau = t_2 - t_1$ represents the temporal thickness (duration) of the hyperbolic temporal slab. Because the medium returns to its original state, the wave frequency recovers to the incident frequency $\omega_1$. By enforcing the continuity of D and B across the temporal boundary at $t = t_2$:

$$B_{z2}(t_2^-) = B_{z3}(t_2^+) \qquad (10.1)$$

$$D_{x2}(t_2^-) = D_{x3}(t_2^+) \qquad (10.2)$$

The second-order transmission and reflection coefficients are analytically derived as:

$$T_2 = 1/2\,(1 + \omega_1/\omega_2) \qquad (11.1)$$

$$R_2 = 1/2\,(1 - \omega_1/\omega_2) \qquad (11.2)$$

## 2. Temporal Transmission and Reflection Coefficients at the Critical angle

### 2.1 Scattering at the First Temporal Boundary

When the plane wave is incident precisely at the critical angle $\theta = \theta_0$, the converted frequency drops to zero ($\omega_2 = 0$). For the region $t > t_1$ the magnetic field ceases to propagate and becomes spatially "frozen" as a constant over time:

$$\boldsymbol{H_2}(t) = \boldsymbol{H_1}(t_1) = const \qquad (12)$$

According to Maxwell's curl equation for the magnetic field ($\nabla \times \boldsymbol{H} = \frac{\partial \boldsymbol{D}}{\partial t}$), it follows that the time derivative of the electric displacement field D must be a non-zero constant:

$$\frac{\partial \boldsymbol{D_2}}{\partial t} = \nabla \times \boldsymbol{H_2} = i\boldsymbol{k} \times \boldsymbol{H_1}(t_1) = const \qquad (13)$$

Integrating this expression with respect to time from the boundary $t_1$ indicates that:

$$\boldsymbol{D_2}(t) = \boldsymbol{D_1}(t_1) + i\big(\boldsymbol{k} \times \boldsymbol{H_1}(t_1)\big)(t - t_1) \tag{14}$$

Using the initial plane wave relationship $\boldsymbol{D_1}(\boldsymbol{t}) = -\frac{\mathbf{1}}{\omega_1}\boldsymbol{k} \times \boldsymbol{H_1}(\boldsymbol{t})$, this equation elegantly simplifies to:

$$\boldsymbol{D_2}(t) = \boldsymbol{D_1}(t_1)\big(1 - i\omega_1(t - t_1)\big) \tag{15}$$

This analytical result demonstrates a profound physical phenomenon: for $t > t_1$, while the magnetic field remains perfectly static, the displacement field **D** (and consequently the electric field **E**) must increase linearly with time to satisfy the fundamental electrodynamic constraints.

### 2.2 Scattering at the Second Temporal Boundary

At the second temporal boundary ($t = t_2$), the evolving fields are scattered into final time-refracted (forward-propagating) and time-reflected (backward-propagating) waves. According to the temporal continuity conditions, the total fields must remain continuous across the boundary:

$$H_3^T(t_2^+) + H_3^R(t_2^+) = H_2(t_2^-) = H_1(t_1^-) \tag{16}$$

where $H_3^T$ and $H_3^R$ represent the magnetic field amplitudes of the final time-refracted and time-reflected waves, respectively. As these two waves correspond to opposite frequencies, time-refracted wave satisfy $D_3^T \propto H_3^T$ and time-reflected wave satisfy $D_3^R \propto -H_3^R$. Applying the continuity condition for the electric displacement field D at $t = t_2$ yields:

$$H_3^T(t_2^+) - H_3^R(t_2^+) = H_1(t_1^-)\big(1 - i\omega_1(t_2 - t_1)\big). \tag{17}$$

Solving the equations provides the field amplitudes:

$$H_3^T(t_2^+) = \left(1 - \frac{i}{2}\omega_1(t_2 - t_1)\right)H_1(t_1^-) \tag{18.1}$$

and

$$H_3^R(t_2^+) = \frac{i}{2}\omega_1(t_2 - t_1)H_1(t_1^-). \tag{18.2}$$

Letting $\tau = t_2 - t_1$ denote the temporal thickness of the hyperbolic slab, the overall transmission ($T_C$) and reflection ($R_C$) coefficients for this critical angle are defined as:

$$T_C = 1 - \frac{i}{2}\omega_1\tau \tag{19.1}$$

and

$$R_C = \frac{i}{2}\omega_1\tau \tag{19.2}$$

# 3. k-Space Evolution of the P-Polarized Gaussian Pulses

### 3.1 Method and theory

Following the theoretical model adapted from Pacheco-Peña et al. [1], the magnetic field distribution for a p-polarized 2D Gaussian pulse propagating along the $x$-axis in vacuum ($\varepsilon_1 = 1$) can be explicitly expressed as:

$$H_z(x,y,t) = H_0 \cdot e^{i\omega_1(x/c-(t-t_0)} \cdot e^{-\frac{1}{2}\left(\frac{x/c-(t-t_0)}{\sigma_t}\right)^2} \cdot e^{-\frac{1}{2}\left(\frac{y}{w_0/2}\right)^2} \tag{20}$$

where $\sigma_t$ denotes the pulse width, $t_0$ represents pulse delay, $w_0$ is the beam waist and $\omega_1$ is the central angular frequency. This expression describes a 2D Gaussian pulse consisting of a transverse spatial profile and a longitudinally propagating envelope. Since the transverse profile $e^{-\frac{1}{2}\left(\frac{y}{w_0/2}\right)^2}$ remains invariant and is independent of the longitudinal dispersion, our primary concern lies within the propagating envelope along the $x$-axis:

$$H_z(x,t) \propto e^{i\omega_1(x/c-(t-t_0)} \cdot e^{-\frac{1}{2}\left(\frac{x/c-(t-t_0)}{\sigma_t}\right)^2} \tag{21}$$

To analyze the spectral characteristics within the hyperbolic temporal slab, we perform a spatial Fourier transform to map the propagating envelope into momentum space:

$$H_k(k,t) \propto \int e^{i\omega_1(x/c-(t-t_0))} \cdot e^{-\frac{1}{2}\left(\frac{x/c-(t-t_0)}{\sigma_t}\right)^2} e^{-ikx}\, dx \tag{22}$$

Assuming $u = x/c - (t - t_0)$, the integral can be evaluated as:

$$H_k(k,t) \propto ce^{-ikc(t-t_0)} \int e^{-\frac{1}{2}\left(\frac{u}{\sigma_t}\right)^2} e^{iu(\omega_1-kc)}\, du \propto c\sigma_t e^{-ikc(t-t_0)} e^{-\frac{\sigma_t^2(\omega_1-kc)^2}{2}} \tag{23}$$

Upon encountering the first temporal boundary at $t = t_1$, the pulse enters the non-dispersive anisotropic hyperbolic medium. The resulting time-refracted (forward-propagating) and time-reflected (backward-propagating) components in $k$-space are respectively given by:

$$H_k^T(k, t > t_1) \propto T_1 e^{-\frac{\sigma_t^2(\omega_1-kc)^2}{2}} e^{\frac{kc}{\sqrt{|\varepsilon_y|}}(t-t_1)} \tag{24.1}$$

$$H_k^R(k, t > t_1) \propto R_1 e^{-\frac{\sigma_t^2(\omega_1-kc)^2}{2}} e^{-\frac{kc}{\sqrt{|\varepsilon_y|}}(t-t_1)} \tag{24.2}$$

where $T_1$ and $R_1$ denotes the first-order temporal transmission and reflection coefficients, and $\frac{kc}{\sqrt{|\varepsilon_y|}}$ represents the modulus of the converted purely imaginary frequency $\omega_2$ (for a pulse propagating along the $x$-axis).

Given that the time-refracted wave experiences exponential growth and rapidly dominates the system inside the temporal slab, the total field spectrum in $k$-space can be tightly approximated by the forward-propagating component. By completing the square within the exponential terms, the total $k$-space spectrum can be rewritten as:

$$H_k(k, t > t_1) \propto T_1 e^{-\frac{\sigma_t^2\left(\omega_1 - kc + \frac{t-t_1}{\sigma_t^2\sqrt{|\varepsilon_y|}}\right)^2}{2}} e^{\frac{\omega_1}{\sqrt{|\varepsilon_y|}}(t-t_1) + \frac{(t-t_1)^2}{2\sigma_t^2|\varepsilon_y|}} \tag{25}$$

This reconstructed expression reveals two profound physical insights. First, the peak of the wavevector spectrum shifts dynamically with time. At the second temporal boundary, the normalized shift of the dominant spectral peak is explicitly derived as:

$$\Delta = \frac{\tau}{\sigma_t^2\sqrt{|\varepsilon_y|}\omega_1} \tag{26}$$

which is proportional to the slab duration $\tau$ and inversely proportional to the square of the pulse width $\sigma_t$. Second, the overall peak amplification factor achieved by the pulse within the hyperbolic temporal slab is quantified as:

$$A = e^{\frac{\omega_1 \tau}{\sqrt{|\varepsilon_y|}} + \frac{\tau^2}{2\sigma_t^2|\varepsilon_y|}} \tag{27}$$

## 3.2 FDTD simulation

To numerically verify the polarization-selective temporal dynamics predicted in the article, we implemented a two-dimensional UPML-FDTD model with a total-field/scattered-field (TF/SF) excitation. The computational window is $40 \times 40\,\mu m^2$ and is discretized by $1000 \times 1000$ uniform cells, corresponding to $\Delta x = \Delta y = 40.04\,nm$. A 60-cell perfectly matched layer (PML) surrounds the domain. The incident field is a temporally modulated Gaussian pulse with carrier wavelength $\lambda_0 = 1\,\mu m$ ($f_1 \approx 299.8THz$, $T \approx 3.336fs$), a pulse width of $\sigma_t = T$, and a transverse Gaussian profile characterized by $w_0 = 3\mu m$. The source propagates predominantly along the $+x$ direction. The FDTD time step is $\Delta t \approx 0.084996fs$, corresponding to 0.9 of the two-dimensional CFL limit.

At $t_1 = 20T \approx 66.712fs$, the background relative permittivity tensor of the central material region is switched abruptly from vacuum ($\varepsilon_1 = 1$) to a hyperbolic material $\varepsilon_2 = diag(2, -5, 2)$ for a duration $\tau$, after which it reverts to $\varepsilon_3 = 1$ at $t_2 = t_1 + \tau$. Two temporal slab durations are considered $\tau_1 = 10\Delta t \approx 0.850fs$ and $\tau_2 = 30\Delta t \approx 2.550fs$. Under a brief interval of $\tau_1$, the analytical shift of the dominant spectral peak ($\Delta_1 \approx 0.0181$) aligns excellently with the simulation result of 0.0183 (shown in Fig. 4C). For a longer interval of $\tau_2$, the analytical value is $\Delta_2 \approx 0.0544$ compared to the simulated 0.0548 (shown in Fig. 4F). This moderate deviation at larger $\tau$ stems from the exponential growth of electromagnetic energy within the temporal slab, which disrupts the ideal absorption conditions of the static PMLs and causes minor non-physical reflections at the computational boundaries.